\documentclass[journal]{IEEEtran}
\usepackage{tikz}
\usepackage[cmex10,intlimits]{amsmath}
\usepackage{amsfonts}
\usepackage{amssymb}
\usepackage{booktabs}
\usepackage[normalem]{ulem}

\usepackage[Symbol]{upgreek}
\usepackage[greek,english]{babel}

\renewcommand{\pi}{\textrm{\greektext p}}
\usepackage[colorlinks]{hyperref}
\hypersetup{
	colorlinks=true, 
	linkcolor=blue!70!black, 
	citecolor=red!70!black, 
	filecolor=blue!70!black, 
	urlcolor=blue!70!black, 
}

\RequirePackage{xcolor} 

\providecommand{\J}{\ensuremath{\mathrm{j}}}        

\providecommand{\D}{\,\mathrm{d}}               
\providecommand{\V}[1]{\boldsymbol{#1}}         
\providecommand{\M}[1]{\mathbf{#1}}             
\providecommand{\T}[1]{\mathrm{#1}}             
\providecommand{\UV}[1]{\hat{\V{#1}}}           

\providecommand{\herm}{\mathrm{H}} 
\providecommand{\trans}{\mathrm{T}}

\providecommand{\ZVAC}{\ensuremath{Z_0}}           

\providecommand{\Ivec}{\ensuremath{\M{I}}}
\providecommand{\Vvec}{\ensuremath{\M{V}}}
\providecommand{\Zmat}{\ensuremath{\M{Z}}}

\newcommand{\ie}{\textit{i}.\textit{e}.{}} 
\newcommand{\eg}{\textit{e}.\textit{g}.{}}
\newcommand{\cf}{\textit{cf}.{}}

\graphicspath{figures/}

\providecommand{\nS}{\tilde{\M{S}}} 
\providecommand{\FFop}{\M{K}}
\providecommand{\nF}{\tilde{\M{F}}}
\providecommand{\bfMoM}{\V{\psi}}
\providecommand{\bfSW}{\V{\phi}}
\providecommand{\weightM}{\V{\Lambda}}

\newcommand{\hfssMark}{\protect\tikz \protect\draw[fill=black!10] (0,0) circle (2pt);}
\newcommand{\comsolMark}{\protect\tikz \protect\draw[fill=black!10,scale=0.05] (0,0) -- (-1,1.5) -- (0,3) -- (1,1.5) -- (0,0);}

\begin{document}

\title{Characteristic Mode Decomposition Using the Scattering Dyadic in Arbitrary Full-Wave Solvers}
\author{Miloslav Capek, \IEEEmembership{Senior Member, IEEE}, Johan Lundgren, \IEEEmembership{Member, IEEE}, Mats Gustafsson, \IEEEmembership{Senior Member, IEEE}, Kurt Schab, \IEEEmembership{Member, IEEE}, and Lukas Jelinek 
\thanks{Manuscript received \today; revised \today. This work was supported by the Czech Science Foundation under project~\mbox{No.~21-19025M} and the Swedish Research Council (2017-04656).}
\thanks{M. Capek and L. Jelinek are with Czech Technical University in Prague, Czech Republic (e-mails: \{miloslav.capek,lukas.jelinek\}@fel.cvut.cz).}
\thanks{J. Lundgren and M. Gustafsson are with Lund University, Lund, Sweden, (e-mails: \{johan.lundgren, mats.gustafsson\}@eit.lth.se).}
\thanks{K. Schab is with the Santa Clara University, Santa Clara, USA (e-mail: kschab@scu.edu).}
\thanks{Color versions of one or more of the figures in this paper are
available online at http://ieeexplore.ieee.org.}
}

\maketitle

\begin{abstract}
Characteristic modes are formulated using the scattering dyadic, which maps incident plane waves to scattered far fields generated by an object of arbitrary material composition. Numerical construction of the scattering dyadic using arbitrary full-wave electromagnetic solvers is demonstrated in examples involving a variety of dielectric and magnetic materials.  Wrapper functions for computing characteristic modes in method-of-moments, finite-difference time domain, and finite element solvers are provided as supplementary material.
\end{abstract}


\section{Introduction}

\IEEEPARstart{C}{haracteristic} mode analysis is a popular tool used in antenna design and the analysis of radiation mechanisms~\cite{MartaEva_TheTCMRevisited, VogelEtAl_CManalysis_PuttingPhysicsBackIntoSimulation,lau2022characteristic}. The most common numerical implementation of characteristic modes is via method-of-moments (MoM) system matrices as proposed by Harrington and Mautz~\cite{HarringtonMautz_TheoryOfCharacteristicModesForConductingBodies}. As such, the characteristic mode decomposition was understood to be closely related -- if not directly depending on~\cite{SarkarMokoleSalazarPalma_AnExposeOnInternalResonancesCM} -- MoM~\cite{Harrington_FieldComputationByMoM}, a procedure for constructing system matrices out of prescribed integro-differential operators~\cite{ChewTongHu_IntegralEquationMethodsForElectromagneticAndElasticWaves}. That this is not necessarily the case was recently presented in~\cite{Gustafsson_etal2021_CMAT_Part1}, where an algebraic link between the transition matrix~\cite{Waterman1965,Kristensson_ScatteringBook} and an arbitrary system matrix was devised, referring back to the original, almost forgotten, definition of characteristic modes given by Garbacz which is based on the perturbation matrix~\cite{Garbacz_1965_TCM}.

A close look at existing eigenvalue problems producing characteristic fields~\cite{GarbaczTurpin_AGeneralizedExpansionForRadiatedAndScatteredFields} and currents~\cite{HarringtonMautz_ComputationOfCharacteristicModesForConductingBodies} reveals that these scattering- and impedance-based formulations represent the same physical problem, only using different representations. This raises an important question of whether other representations exist and, if so, whether they have advantageous properties over existing approaches. This paper addresses these questions utilizing a scattering dyadic~\cite{Kristensson_ScatteringBook}, which affords the possibility for characteristic modes to be evaluated in arbitrary numerical methods, to be calculated for objects constructed from arbitrary materials, and to be reconstructed from far-field measured data.

An object's scattering dyadic transforms incident plane waves to scattered far fields. In this paper, it is shown that characteristic modes can be calculated directly from a matrix representation of the scattering dyadic that may be constructed from only the solution of a series of plane wave scattering problems. Hence, the finite element method (FEM)~\cite{Jin_FEM2014} or finite-difference time-domain method (FDTD)~\cite{Taflove_2004_FDTD} may be utilized, giving the opportunity to deal with arbitrary distributions of anisotropic dielectric and magnetic materials. An additional advantage is that the unknowns used in the underlying numerical method are decoupled from the degrees of freedom to express characteristic modes, \ie{}, a studied object can be described and efficiently evaluated with millions of unknowns but the matrix representing the scattering dyadic used for characteristic mode eigenvalue decomposition can be significantly smaller, \eg{}, hundreds of degrees of freedom for practical problems.

The proposed method is versatile, simple to implement, yet general. It directly produces characteristic fields, which can be used for powerful far-field tracking as proposed in~\cite{Gustafsson_etal2021_CMAT_Part2}. Furthermore, utilization of FDTD makes it possible to efficiently recover broadband characteristic mode data. Finally, since a discrete form of the scattering dyadic can also be assembled from bi-static measurements, there is a possibility to reconstruct modal properties of realized antennas and scatterers.

The paper is organized as follows. Section~\ref{sec:sd-modes} describes the formulation of characteristic modes using the scattering dyadic along with high-level connections to formulations involving the transition matrix~\cite{Gustafsson_etal2021_CMAT_Part1} and system (impedance) matrices~\cite{HarringtonMautz_ComputationOfCharacteristicModesForConductingBodies}.  In Section~\ref{sec:numerical}, practical aspects of collecting necessary scattering dyadic data using MoM, FEM, and FDTD are discussed.  In Section~\ref{sec:examples}, example calculations involving PEC structures as well as those involving dielectric and magnetic materials are used to demonstrate the method's efficacy and to draw general conclusions about the application of different solvers in calculating characteristic modes. Important aspects of the method are discussed in Section~\ref{sec:disc}.  Section~\ref{sec:concl} concludes the paper with discussion points and open problems. The paper is accompanied by supplementary material containing wrappers for several contemporary EM simulation packages (CST Studio Suite~\cite{CST}, Altair FEKO~\cite{Altair2022}, Ansys HFSS~\cite{hfss2021}, COMSOL Multiphysics~\cite{comsol}).

\section{Modal Decomposition of Scattering Dyadic}
\label{sec:sd-modes}

\begin{figure}
    \centering
    \includegraphics[width=\columnwidth]{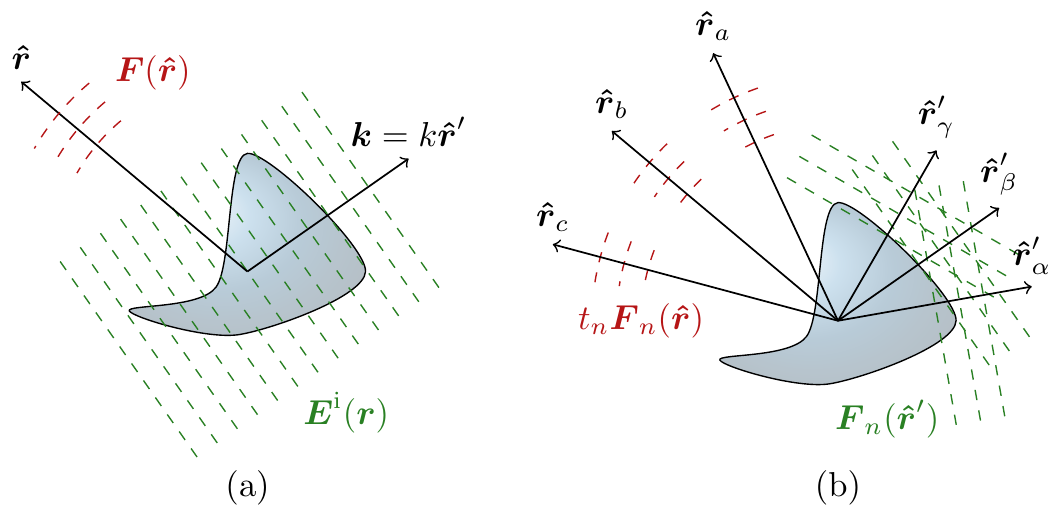}
    \caption{(a) Description of an obstacle via scattering dyadic~$\V{S}\left(\UV{r},\UV{r}'\right)$ where the object is illuminated by an incident plane  wave~$\V{E}^\T{i}$ with vector amplitude~$\V{E}_0$ propagating in the direction~$\UV{r}'$ producing scattered far field~$\V{E}^\T{s}$ with vector amplitude $\V{F}$ measured in direction~$\UV{r}$. (b) Characteristic modes are incident plane wave spectra described by~$\V{F}_n$ which produce scattered scaled far fields with the same pattern, \ie{}, $t_n\V{F}_n$.  In both panels, dashed lines schematically represent wavefronts of the incident plane (green) and scattered spherical (red) waves.}
    \label{fig:schematics}
\end{figure}

Consider an object illuminated by an incident electric field~$\V{E}^\T{i}$ and the resulting scattered field~$\V{E}^\T{s}$.  Let the electric far field $\V{F}$ scattered by the object be defined as~\cite{Balanis1989},
\begin{equation}
   \V{E}^\T{s}(\V{r}) \approx  \V{F} (\UV{r})\frac{\T{e}^{-\J k r}}{r},\quad \text{as}~r\rightarrow\infty,
    \label{eq:ff-def}
\end{equation}
with the radial unit vector~$\UV{r}$, radial vector~$\V{r} = r \UV{r}$, wavenumber~$k$, and $\J$~being the imaginary unit. The far-field scattering properties of the obstacle are fully characterized by a scattering dyadic $\V{S}$, which maps the vector amplitude~$\V{E}_0$ of an incident plane wave propagating in the~$\UV{r}'$ direction into the far field~$\V{F}$ in the $\UV{r}$ direction, see Fig.~\ref{fig:schematics}a, \ie{},
\begin{equation}
    \V{F}(\UV{r})=  \J \dfrac{4\pi}{k} \V{S}(\UV{r},\UV{r}') \cdot \V{E}_0 (\UV{r}'),
    \label{eq:scattering-dyadic-def}
\end{equation}
where the incident field is assumed to take the form
\begin{equation}
    \V{E}^\T{i} \left(\V{r} \right) = \V{E}_0 (\UV{r}') \, \T{e}^{-\J k \hat{\V{r}}' \cdot \V{r}}.
    \label{eq:ei-pw}
\end{equation}
Here the scattering dyadic $\V{S} (\UV{r},\UV{r}')$ contains a non-standard scaling factor $-\J k/ (4 \pi)$, \cf{} \cite[(4.29)]{Kristensson_ScatteringBook}, for the sake of notational and dimensional simplicity in later sections. Note that the case of $\UV{r} = \UV{r}'$ corresponds to forward scattering and that reciprocal systems exhibit the symmetry property 
\begin{equation}
    \V{S}(\UV{r},\UV{r}') = \V{S}^\trans(-\UV{r}',-\UV{r}).
\end{equation} 
For excitations consisting of a continuum of plane-waves~\cite{Kristensson_ScatteringBook}, with propagation directions covering the unit sphere, the scattered far field is similarly given by
\begin{equation}
    \V{F}(\UV{r}) = \J \dfrac{4\pi}{k} \int \limits_{4\pi} \V{S}(\UV{r},\UV{r}') \cdot \V{E}_0 (\UV{r}') \D{\varOmega'},
\label{eq:scatDyad:analytic}
\end{equation}
where the integration limit $4\pi$ denotes integration of~$\UV{r}'$ over the unit sphere. Notice that to keep the notation simple, the same symbol~$\V{E}_0 (\UV{r}')$ is used in~\eqref{eq:scattering-dyadic-def} and~\eqref{eq:scatDyad:analytic} although their physical meaning is slightly different. In~\eqref{eq:scatDyad:analytic} the meaning is altered to the angular density with unit~$\T{V \T{m}^{-1} \T{sr}^{-1}}$ .

By \eqref{eq:scatDyad:analytic}, the scattering dyadic transforms an incident spectrum of plane waves described by the vector function $\V{E}_0(\UV{r}')$ into the scattered far field $\V{F}(\UV{r})$ analogously to how a transition matrix transforms regular spherical waves into outgoing spherical waves~\cite{Kristensson_ScatteringBook} or an admittance matrix transforms incident electric field into induced electric currents~\cite{Harrington_FieldComputationByMoM}. As such, an eigenvalue decomposition of the scattering dyadic can be used to generate eigenfunctions~$\V{F}_n$, which, when applied as incident fields, produce scattered fields with the same pattern, \ie{},
\begin{equation}
\int \limits_{4\pi} \V{S}(\UV{r},\UV{r}') \cdot \V{F}_n(\UV{r}') \D{\varOmega}' =  t_n \V{F}_n(\UV{r}).
\label{eq:cm-s}
\end{equation}
As will be discussed in the following subsections and Appendix~\ref{app:StoZ}, these eigensolutions are alternative representations of characteristic modes obtained through transition~\cite{Gustafsson_etal2021_CMAT_Part1} and impedance matrix~\cite{HarringtonMautz_TheoryOfCharacteristicModesForConductingBodies} methods. In \eqref{eq:cm-s}, the eigenvalues $t_n$ describing the relative scaling of incident and scattered fields. This concept is illustrated in Fig.~\ref{fig:schematics}b, where an object is illuminated by an eigenexcitation consisting of spectrum of plane waves with vector amplitudes $\V{F}_n(\UV{r}')$ and the scattered field at all locations on the far-field sphere is a scaled copy of the incident plane wave spectrum, \ie{}, $t_n\V{F}_n(\UV{r})$.  

\subsection{Relation to the Transition Matrix Decomposition}
\label{sec:equiv-g}

The eigenvalue problem in~\eqref{eq:cm-s} closely resembles characteristic modes as defined by the transition matrix \cite{Garbacz_TCMdissertation, Gustafsson_etal2021_CMAT_Part2}.  In that definition, characteristic modes are produced by the eigenvalue problem
\begin{equation}
    \M{T} \M{f}_n = t_n \M{f}_n,
    \label{eq:TCMA}
\end{equation}
where the vectors~$\M{f}_n$ represent spherical wave expansion coefficients and the transition matrix~$\M{T}$ maps regular spherical waves onto outgoing spherical waves~\cite{Kristensson_ScatteringBook}. The equivalence of eigenvalue problems~\eqref{eq:cm-s} and~\eqref{eq:TCMA} is detailed Appendix~\ref{app:TtoS} and is based on a relation between the transition matrix and scattering dyadic
\begin{equation}
T_{\alpha \beta} = \int \limits_{4\pi} \int \limits_{4\pi} \bfSW_\alpha^\ast (\UV{r}) \cdot \V{S}(\UV{r},\UV{r}') \cdot \bfSW_\beta (\UV{r}') \D{\varOmega}\D{\varOmega}',
\label{eq:TSrel}
\end{equation}
where~$\left\{\bfSW_\alpha\right\}$ is a basis of spherical harmonics, see Appendix~\ref{app:TtoS}. This basis also connects modal far fields~$\V{F}_n(\UV{r})$ with the eigenvectors~$\M{f}_n$ via
\begin{equation}
\V{F}_n(\UV{r}) = \sqrt{\ZVAC} \sum_\alpha f_{n,\alpha} \bfSW_\alpha(\UV{r})
\end{equation}
with $\ZVAC$ being the free-space impedance.

\subsection{Relation to the Impedance Matrix Decomposition}
\label{sec:equiv-hm}

The more common method of generating the characteristic modes of a lossless scatterer considers the generalized eigenvalue problem~\cite{HarringtonMautz_ComputationOfCharacteristicModesForConductingBodies}
\begin{equation}
	\M{X}\M{I}_n = \lambda_n\M{R}\M{I}_n,
\label{eq:CMAZ}
\end{equation}
where $\M{X} = \T{Im}\{\M{Z}\}$ and $\M{R} = \T{Re}\{\M{Z}\}$ are the reactance and radiation components, respectively, of impedance matrix~$\M{Z}$ evaluated via a method-of-moments formulation of the electric field integral equation\footnote{Formula~\eqref{eq:CMAZ} can be applied for other formulations as well, identifying proper reactance and radiating operators~\cite{Gustafsson_etal2021_CMAT_Part1}.}~\cite{Harrington_TimeHarmonicElmagField}, $\M{I}_n$~are vectors of expansion coefficients representing characteristic current densities~$\V{J}_n$, and $\lambda_n$~are the eigenvalues related to~$t_n$ via \cite{Gustafsson_etal2021_CMAT_Part1}
\begin{equation}
t_n = -\frac{1}{1 + \J \lambda_n}.
\label{eq:tn2lamn}
\end{equation}

The connection between~\eqref{eq:cm-s} and~\eqref{eq:CMAZ} is shown in Appendix~\ref{app:StoZ} utilizing the scattering dyadic formulation~\eqref{eq:cm-s} as a starting point. Appendix~\ref{app:StoZ} also justifies the extraction of the elements of the scattering dyadic directly from the impedance matrix via
\begin{equation}
S_{\gamma\gamma'}(\UV{r},\UV{r}') = -\dfrac{1}{\ZVAC} \FFop_\gamma(\UV{r})\Zmat^{-1}\FFop_{\gamma'}^{\herm}(\UV{r}'),
\label{eq:SfromZ}
\end{equation}
where the matrix~$\FFop_\gamma(\UV{r})$ transforms a current vector~$\M{I}$ into its $\gamma$-polarized far field defined by~\eqref{eq:ff-def} as~\cite{GustafssonTayliEhrenborgEtAl_AntennaCurrentOptimizationUsingMatlabAndCVX}
\begin{equation}
F_{\gamma} \left(\hat{\V{r}} \right) = \FFop_\gamma \left(\hat{\V{r}} \right) \M{I}
\label{eq:FFfromI}
\end{equation}
with polar and azimuthal components $\gamma = \left\{ \vartheta, \varphi \right\}$.

Formula~\eqref{eq:SfromZ} prescribes how to evaluate scattering dyadic with method-of-moments solver. As such, it is an important side-product of this paper, and, together with~\eqref{eq:tn2lamn}, serves as independent validation of equality between~\eqref{eq:cm-s} and~\eqref{eq:CMAZ}. It is also worth mentioning that \eqref{eq:SfromZ} defines the far-fields scattered by currents excited by plane waves and that these currents might be further processed to form a set of so-called characteristic basis functions~\cite{PrakashMittra_CHBFM} which are sometimes employed to reduce the order of electromagnetic models.

\subsection{Evaluation of Characteristic Quantities}
\label{sec:characteristic-quantities}

The eigenvectors~$\V{F}_n$ in \eqref{eq:cm-s} represent both characteristic scattered far fields and characteristic excitations, \ie{}, it is possible to extract all (modal) metrics by applying
\begin{equation}
    \V{E}_n^\T{i}(\V{r}) = -\J \dfrac{k}{4 \pi t_n}\int \limits_{4\pi} \V{F}_n(\UV{r}') \T{e}^{-\J k\UV{r}'\cdot\V{r}}\D{\varOmega'}
    \label{eq:cm-en}
\end{equation}
as a characteristic incident field impinging on the scatterer using any method of solving Maxwell's equations.  

For example, adopting~\eqref{eq:cm-en}, to obtain the modal current density within method of moments reads
\begin{equation}
\Ivec_n = \Zmat^{-1} \Vvec_n,
\label{eq:InFromVn}
\end{equation}
where the excitation is constructed using
\begin{equation}
V_{n,i} = \int_{\mathbb{R}^3} \bfMoM_i (\V{r}) \cdot \V{E}_n^\T{i} (\V{r}) \D{V}
\label{eq:VnFromEnI}
\end{equation}
and the incident field defined in~\eqref{eq:cm-en}. Hence, once the characteristic mode excitation~$\V{F}_n$ is known, one additional run of the chosen solver must be executed to extract each characteristic mode's field and current distributions\footnote{If induced field and current data are stored for each incident plane wave during the construction of the scattering dyadic, these additional solutions may be avoided, see Section~\ref{sec:numerical}.}.  Similar relations may be used to extract modal quantities from scattering-based characteristic modes calculated in non-MoM solvers (\eg{}, FDTD, FEM) as discussed in Section~\ref{sec:numerical}.

Characteristic current distributions are most commonly the quantities of interest due to the equivalencies described in Sections~\ref{sec:equiv-g} and \ref{sec:equiv-hm} and the prevalence of MoM-based characteristic mode formulations.  However, examining induced current distributions is not strictly required for characteristic mode analysis and, depending on the chosen approach to solving Maxwell's equations, the study of other characteristic quantities, such as characteristic near fields, may be advantageous. The extraction of these modal quantities follows the same procedure described above by studying the metric of interest produced by characteristic excitations.

\section{Numerical Evaluation of Scattering Dyadic}
\label{sec:numerical}

\begin{figure}[t]
    \centering
    \includegraphics[width=\columnwidth]{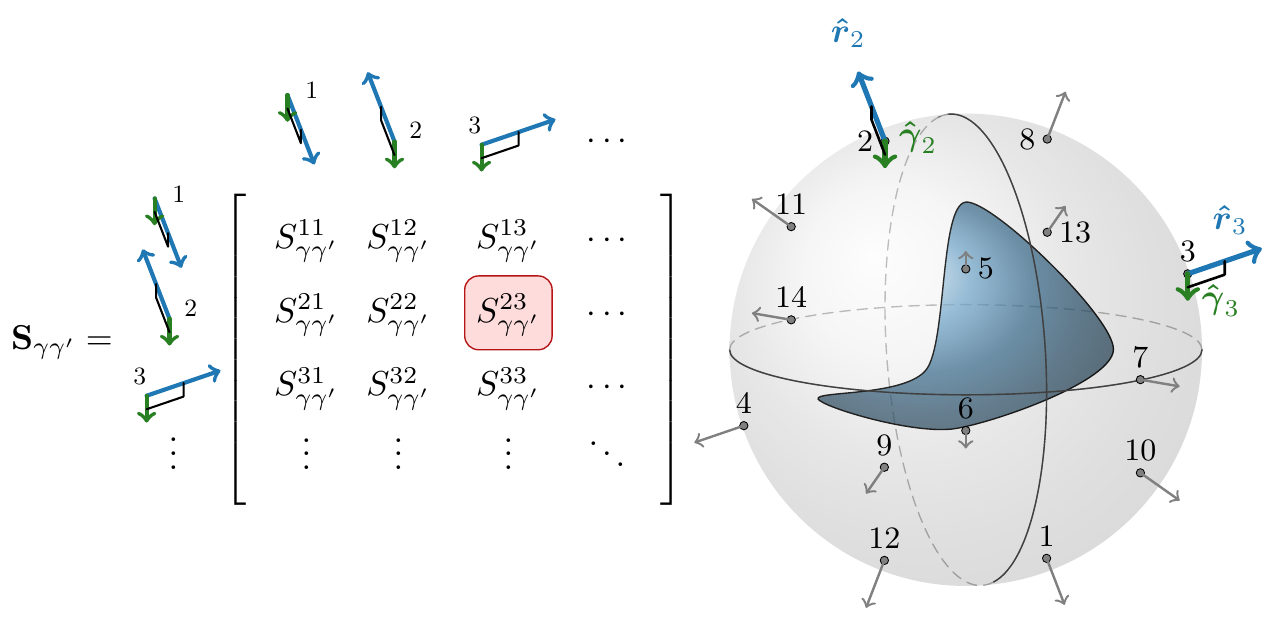}
    \caption{Construction of the matrix representation~$\M{S}$ of scattering dyadic as far field measured in direction~$\UV{r}_p$ with polarization~$\UV{\gamma}_p$ as a reaction to plane wave excitation propagating in the direction~$\UV{r}_q$ with polarization~$\UV{\gamma}_q$. The matrix element for $p = 2$ and $q = 3$ is highlighted. Nystr\"{o}m method transforms integral formula~\eqref{eq:scatDyad:analytic} to its algebraic form, \ie{}, each $q$-th element of the matrix is weighted by quadrature weight~$l_q$. The number of quadrature points is~$N_\T{q} = 14$}.
    \label{fig:schematics2}
\end{figure}

A finite-dimensional representation of the eigenvalue problem in~\eqref{eq:cm-s} may be constructed by applying the Nystr\"{o}m method~\cite{TongChew_NystromMethod_2020}, approximating the integration over the unit sphere by a chosen quadrature rule consisting of $N_\T{q}$ sample locations~$\UV{r}_q'$ and weights~$\ell_q$ for $q\in\left\{1,\dots,N_{\T{q}}\right\}$, see Fig.~\ref{fig:schematics2}. The resulting approximation of the eigenvalue problem~\eqref{eq:cm-s} reads
\begin{equation}
    \sum_{q=1}^{N_\T{q}} \V{S}(\UV{r},\UV{r}_q') \ell_{q} \cdot \V{F}_n(\UV{r}_q') = t_n \V{F}_n(\UV{r}).
    \label{eq:quadrature-s-eig}
\end{equation}
Let the far field at each quadrature sampling point~$\UV{r}'_q$ be expressed as a sum of two transverse vector components
\begin{equation}
    \V{F}(\UV{r}_q') = F^q_{\vartheta'}\UV{\vartheta}{}'+F^q_{\varphi'}\UV{\varphi}{}'
\end{equation}
and denote the elements of the scattering dyadic at pairs of quadrature points as 
\begin{equation}
    \V{S}(\UV{r}_p,\UV{r}{}_q') =
    S_{\vartheta\vartheta'}^{pq}\UV{\vartheta}\UV{\vartheta}{}' + S_{\vartheta\varphi'}^{pq}\UV{\vartheta}\UV{\varphi}' + S_{\varphi\vartheta'}^{pq}\UV{\varphi}\UV{\vartheta}{}' + S_{\varphi\varphi'}^{pq}\UV{\varphi}\UV{\varphi}{}'.
\end{equation}
Evaluating~\eqref{eq:quadrature-s-eig} at all quadrature locations~$\UV{r}_p$ leads to the eigenvalue problem
\begin{equation}
    \nS\nF_n = t_n \nF_n
    \label{eq:cm-s-discrete}
\end{equation}
where the square matrix $\nS$ is organized as
\begin{equation}
\nS = \begin{bmatrix}
    \M{S}_{\vartheta\vartheta'} & \M{S}_{\vartheta\varphi'} \\
    \M{S}_{\varphi\vartheta'} & \M{S}_{\varphi\varphi'} \end{bmatrix}
\begin{bmatrix}
    \weightM & \M{0}\\
    \M{0} & \weightM \end{bmatrix}
\end{equation}
where the weighting matrix~$\weightM$ is a diagonal matrix with elements~$\ell_q$ and the eigenvectors $\nF_{n}$ are organized as 
\begin{equation}
    \nF_n = \begin{bmatrix}
    \M{F}_{n,\vartheta}\\
    \M{F}_{n,\varphi}
    \end{bmatrix},
\end{equation}
where the vectors $\M{F}_{n,\gamma}$ contain far-field coefficients $F^p_{n,\gamma}$ of each mode collected at all quadrature points. By~\eqref{eq:scattering-dyadic-def}, the interpretation of each element of the matrix $\M{S}_{\gamma\gamma'} = \left[S^{pq}_{\gamma\gamma'}\right]$ is that of the $\gamma$-polarized far-field amplitude at the observation direction~$\UV{r}_p$ when the scatterer is illuminated with a $\gamma'$-polarized plane wave propagating along the~$\UV{r}'_q$ direction, where~$\gamma\in\left\{\vartheta,\varphi\right\}$ and $\gamma'\in\left\{\vartheta',\varphi'\right\}$ represent the two choices for both incident and observed field polarizations, see Fig.~\ref{fig:schematics2}.

An important parameter of the proposed method is the number of plane waves~$2N_\T{q}$ (counting two polarizations) used to construct the matrix~$\nS$. The parameter~$N_{\T{q}}$ controls the overall precision of this discrete representation of the scattering dyadic. For the particular case of Lebedev quadrature rule~\cite{LebedevLaikov_QuadratureRuleSphere}, a sufficient number of plane waves used to represent the scattering dyadic of a scatterer of electrical size $ka$ can be estimated from
\begin{equation}
N_\T{q} \geq \dfrac{4}{3} \left( ka + 2 \sqrt[3]{ka} + 1 \right)^2,
\label{eq:LebQuadEstimate}
\end{equation}
based on the maximum number of spherical vector waves that can accurately be integrated via this quadrature rule~\cite{WANG2003}. Note that, for Lebedev quadrature, the number of quadrature points $N_\T{q}$ must be selected from a specific set of integers, so the above rule should be understood as taking the lowest member of that set satisfying the inequality. Regardless of the chosen quadrature rule and meshing, the precision of the characteristic mode decomposition~\eqref{eq:cm-s-discrete} for lossless structures can be verified by checking the condition~\cite{Gustafsson_etal2021_CMAT_Part1}
\begin{equation}
\vert 2 t_n + 1 \vert = 1.
\label{eq:convCheck}
\end{equation}

The matrix~$\nS$ can be filled by repeated excitation of a structure with plane waves in directions~$\UV{r}'_q$ described by the selected quadrature rule and observing the scattered fields at locations~$\UV{r}_p$, also based on the chosen quadrature rule.  Because the scattered fields at all selected directions can generally be computed with negligible cost compared to that of solving the underlying plane wave scattering problem, the overall computational cost of populating the matrix~$\nS$ scales linearly with the parameter~$N_\T{q}$. Nevertheless, parallelization can be readily employed in frequency domain solvers for simultaneous evaluation of multiple frequency points and FDTD for simultaneous evaluation of incident plane wave excitations. Frequency domain methods may also be efficiently implemented using direct solvers that reuse matrix inversion or factorization (\eg{} LU decomposition) for multiple plane wave excitations. This approach necessarily becomes infeasible as the system matrix becomes very large and where iterative solutions must be used instead, in which only partial solution reuse may be possible~\cite{Saad2003_IterativeMethods}.

The following subsections outline particular considerations for using three common numerical methods, MoM, FEM, and FDTD, to numerically compute the matrix~$\nS$, as well as a short discussion of applying far-field modal tracking using scattering dyadic data.

\subsection{Method of Moments}
\label{sec:MoM}

Apart from rare exceptions~\cite{hu2016fe,2022_Paschaloudis_FEM_CMA,Gustafsson_etal2021_CMAT_Part2}, MoM is the only method used for characteristic mode decomposition.  Despite the ease of using the system matrix approach to compute characteristic modes using MoM data~\cite{HarringtonMautz_ComputationOfCharacteristicModesForConductingBodies}, MoM can also be utilized in the calculation and decomposition of the scattering dyadic. The advantage of utilizing MoM in the scattering dyadic approach to characteristic modes is that it is independent of which MoM formulation, \eg{}, EFIE, MFIE, CFIE, PMCHWT, or MLFMA~\cite{Jin_TheoryAndComputationOfElectromagneticFields} is used, since only plane-wave scattering problems must be solved and no decomposition of system (impedance) operators is necessary.

When surface equivalence methods are employed, the calculation of the matrix~$\nS$ via MoM excels in precision and computational time. As compared to contemporary approaches~\cite{YlaOijala_PMCHWTBasedCharacteristicModeFormulationsforMaterialBodies, YlaOijala_GeneralizedTCM2019}, the form of the decomposition is always the same, \cf{}~\eqref{eq:cm-s-discrete}, no matter how the equivalence is formulated. For dielectric and magnetic materials, it can be combined with FEM, where MoM serves as a radiation boundary condition~\cite{Jin_TheoryAndComputationOfElectromagneticFields}.

A drawback of MoM lie mainly in dealing with inhomogeneous materials, which is infeasible using surface equivalence principle. Volumetric MoM is well suited to dealing with inhomogeneous media, but its computational cost scales rapidly for electrically large problems, though this can be somewhat mitigated through the use of FFT acceleration~\cite{Jin_TheoryAndComputationOfElectromagneticFields}.

\subsection{Finite-Difference Time-Domain Method}
\label{sec:FDTD}

In FDTD, the scattering dyadic is determined by illumination of the object with time-domain pulses from different directions. Absorbing boundary conditions, \eg{}, perfectly matched layers (PML), are used to mimic free space and near to far-field transformation is used to evaluate the scattered far field~\cite{Jin_TheoryAndComputationOfElectromagneticFields}. Because broadband incident pulses can be used to produce scattering dyadic data over broad bandwidths, FDTD affords the ability to compute broadband characteristic mode data with a number of calculations scaling with the parameter~$N_\T{q}$ (as opposed to $N_\T{q}$ times the number of desired frequency points, as occurs in frequency domain solvers). FDTD is matrix-free and time stepping is easily accelerated by graphic processing units (GPUs)~\cite{Warren2019}.

Drawbacks with FDTD include numerical dispersion producing directional phase shifts in the scattering dyadic, low-order approximation of curved surfaces, increased computational time for narrow resonances due to ringing effects, and limited possibilities to accelerate computations for multiple excitations without the use of parallelization.

\subsection{Finite Element Method}
\label{sec:FEM}

Similarly to MoM, FEM solvers may be used to construct the matrix~$\nS$ by solving plane-wave excitation problems in the frequency domain. FEM methods excel in the study of arbitrarily shaped inhomogeneous scatterers, but similarly to FDTD, the solution domain in FEM calculations must be terminated either by absorbing boundary conditions, \eg{}, PML or by a boundary surface modelled by MoM~\cite{Jin_TheoryAndComputationOfElectromagneticFields}. The latter approach removes the computational burden associated with conventional absorbing boundary conditions and modeling of a truncated free space region surrounding the scatterer. Regardless of which approach is taken, the numerical accuracy of scattering dyadic data produced by FEM is dependent not only on the discrete representation of the scatterer itself but also on the treatment of the open boundary.  In large problems, the rapid growth of the size of FEM system matrix necessitates the use of iterative solvers, which lowers the efficiency of repeated evaluation of scattering from individual plane wave excitations~\cite{Saad2003_IterativeMethods}.

\subsection{Far-Field Tracking}
\label{sec:tracking}

Tracking is a set of post-processing techniques for studying and interpreting the modal quantities across discrete frequency points. Many procedures exist~(see~\cite{capek2022computational} for a detailed reference list), typically based on correlation between eigenvectors at two adjacent frequency points. Since the eigenvectors in~\eqref{eq:cm-s-discrete} are themselves the modal far fields, far-field tracking is, similarly to~\cite{Gustafsson_etal2021_CMAT_Part2}, adopted here using
\begin{equation}
    \max_{n}\left \vert \nF_m^{\herm} (k_q) \begin{bmatrix}
    \weightM & \M{0}\\
    \M{0} & \weightM \end{bmatrix} \nF_n(k_{q+1}) \right \vert,
\label{eq:Ftracking}
\end{equation}
where characteristic far fields are assumed to be normalized to unit radiated power, $k_q$ and $k_{q+1}$ are wavenumbers corresponding to $q$-th and $(q+1)$-th frequency points, and~$m$ denotes the index of a chosen characteristic far field for which the most similar one obtained by searching over the index~$n$. Note that the weighting matrix~$\V{\Lambda}$ is introduced in~\eqref{eq:Ftracking} so that the maximized quantity represents the inner product of modal far fields over the unit sphere.

\section{Examples}
\label{sec:examples}

Here we present examples illustrating the practical calculation of characteristic modes using the scattering dyadic formulation and several numerical methods, namely MoM, FEM, and FDTD, as implemented in CST Studio Suite~\cite{CST}, Altair FEKO~\cite{Altair2022}, Ansys HFSS~\cite{hfss2021}, COMSOL Multiphysics~\cite{comsol}, and AToM~\cite{atom}. Templates and scripts for these solvers may be found in supplementary material~\cite{ScatDyad22_SuplMat}. Examples are selected to demonstrate agreement among differing numerical methods, agreement with analytic results, and the analysis of a practical structure involving both dielectric and metallic materials. Throughout all examples, with an exception of COMSOL Multiphysics, the number of quadrature points $N_\T{q}$ is fixed across electrical size $ka$ and set lower than the conservative estimate given in \eqref{eq:LebQuadEstimate}. Evaluations within the COMSOL environment used~\eqref{eq:LebQuadEstimate} at at every frequency point.
 The selected number of quadrature points differs between implementations in different solvers to reduce the computational cost, though in all cases it is listed next to corresponding results.  A detailed study of accuracy and quadrature settings is presented later as part of the discussion in Section~\ref{sec:disc}.

\subsection{PEC Plate}

\begin{figure}[t]
    \centering
    \includegraphics[width=\columnwidth]{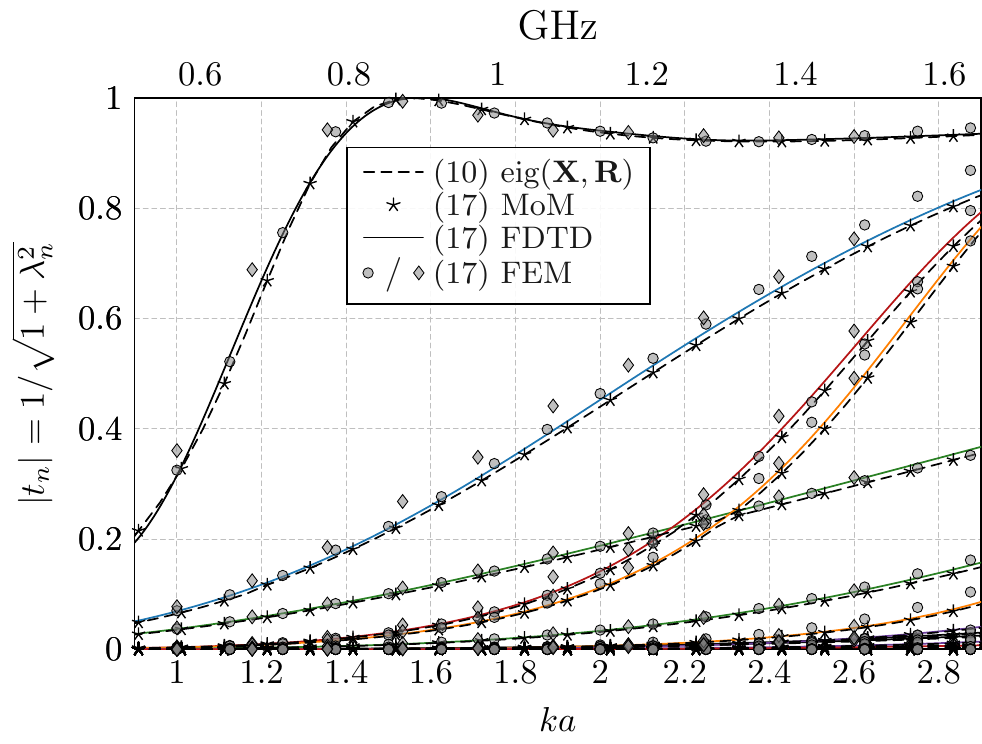}
    \caption{Modal significances~$|t_n|$ of a rectangular PEC plate with dimensions \mbox{$\ell\times\ell/2$} for $\ell=15\,\T{cm}$. Data are shown for calculations using the system matrix formulation \eqref{eq:CMAZ} (AToM, dashed lines) and the scattering dyadic formulation \eqref{eq:cm-s-discrete} using MoM (FEKO, $\star$, $N_\T{q} = 50$), FDTD (CST, solid lines, $N_\T{q} = 110$), and FEM (HFSS, \hfssMark\,, $N_\T{q} = 14$; COMSOL, \comsolMark, $N_\T{q}$ given by~\eqref{eq:LebQuadEstimate}).}
    \label{fig:plate_abst}
\end{figure}

\begin{figure}[t]
    \centering
    \includegraphics[width=\columnwidth]{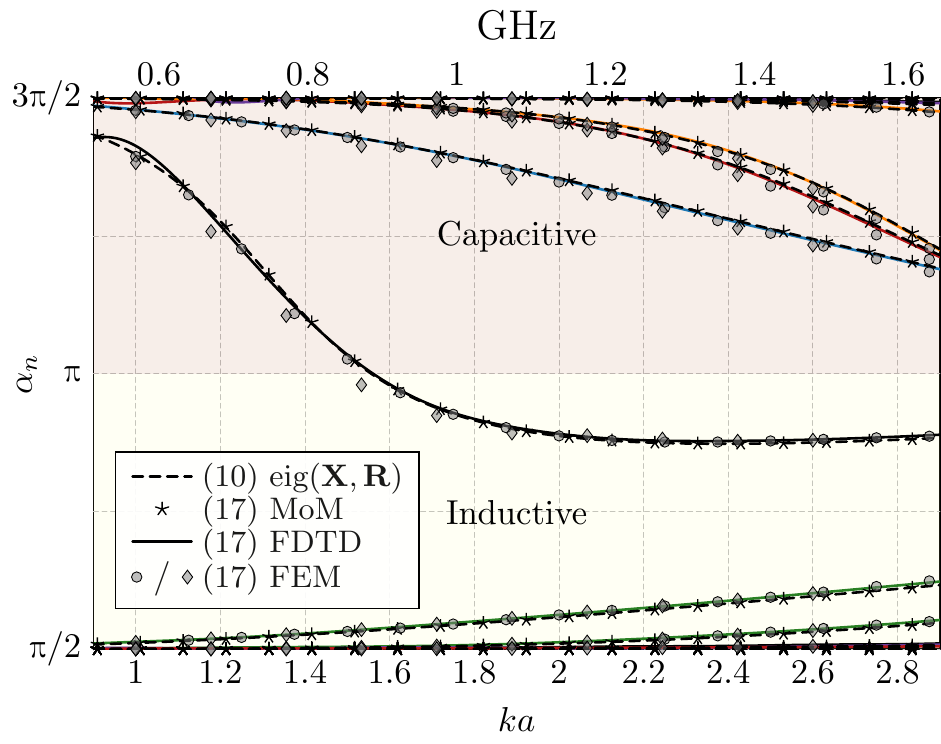}
    \caption{Characteristic angles~$\alpha_n$ of a rectangular PEC plate with dimensions \mbox{$\ell\times\ell/2$} for $\ell=15\,\T{cm}$. Data are shown for calculations using the system matrix formulation \eqref{eq:CMAZ} (AToM, dashed lines) and the scattering dyadic formulation \eqref{eq:cm-s-discrete} using MoM (FEKO, $\star$, $N_\T{q} = 50$), FDTD (CST, solid lines, $N_\T{q} = 110$), and FEM (HFSS, \hfssMark\,, $N_\T{q} = 14$; COMSOL, \comsolMark, $N_\T{q}$ given by~\eqref{eq:LebQuadEstimate}).}
    \label{fig:plate_alpha}
\end{figure}

As a first example, we consider a perfectly electrically conducting (PEC) rectangular plate of dimension $\ell\times\ell/2$ ($\ell=15\,\T{cm}$). Because of the structure's simplicity, it is straightforward to analyze using most full-wave solvers.  Characteristic modes are calculated via \eqref{eq:cm-s-discrete} using matrices~$\nS$ obtained from each method.  Modal significances~$|t_n|$ and characteristic angles\footnote{For modes with low modal significance $|t_n|\ll 1$, the characteristic angle~$\alpha_n$ calculated by \eqref{eq:char-angle} is highly sensitive to small numerical errors, leading to jumps.  These errors can largely be avoided by a slight reformulation $\alpha_n = \arg(1+2 t_n)/2 + \pi/2 \in [\pi/2,3\pi/2]$ and excluding modes with $|t_n|\leq\epsilon$, where the parameter $\epsilon$ effectively sets the maximum detectable eigenvalue magnitude $|\lambda_n| \leq \epsilon^{-1}$ in the system matrix formulation~\eqref{eq:CMAZ}.}~\cite{Gustafsson_etal2021_CMAT_Part1}
\begin{equation}
    \alpha_n = \arg(t_n)\,\in\,\left[\pi/2,3\pi/2\right]
    \label{eq:char-angle}
\end{equation}
from all calculations are shown in Figs.~\ref{fig:plate_abst} and~\ref{fig:plate_alpha}, respectively.  All methods show good agreement in modal significance and characteristic angle, with slight deviation in FEM results at higher frequencies.

\begin{figure}[t]
    \centering
    \includegraphics[width=\columnwidth]{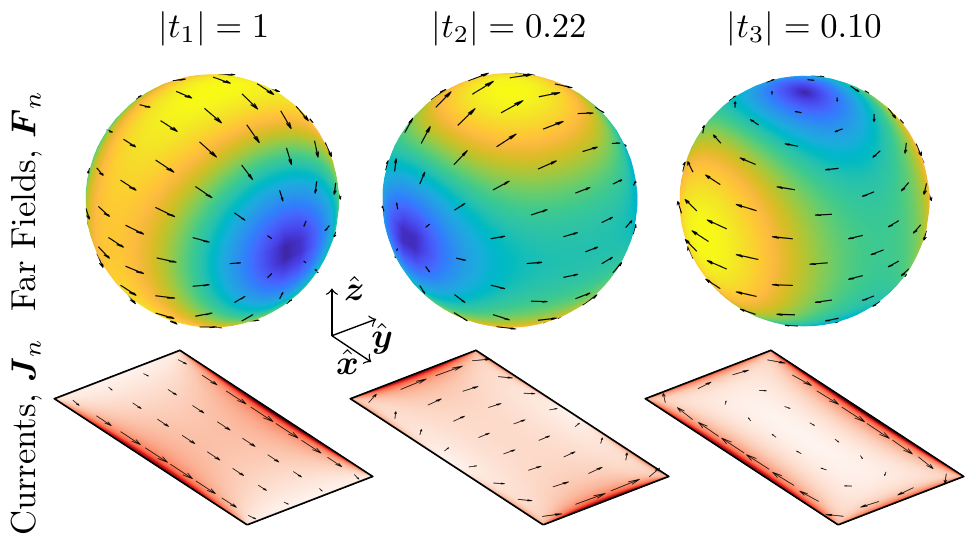}
    \caption{Characteristic far fields~$\V{F}_n(\UV{r})$ (top) and current densities~$\V{J}_n$ (bottom) for the three modes of highest modal significance for a rectangular PEC plate with dimensions \mbox{$\ell\times\ell/2$} for $\ell=15\,\T{cm}$ evaluated at $0.9\,\T{GHz}$ using 110 Lebedev quadrature points in FDTD.  The polarization and amplitude of far fields (top) are displayed at the Lebedev quadrature points by arrows. The colormap displays the far-field amplitude. The corresponding characteristic currents (real valued) on the plate are similarly represented in the bottom panels.}
    \label{fig:plate_quantities}
\end{figure}

Calculation of the eigenvectors~$\nF_n$ and eigenvectors $t_n$ requires only numerical computation of the matrix~$\nS$.  In non-MoM solvers, however, the corresponding characteristic fields or current densities must be obtained by illuminating the structure with multiple plane-wave excitations described by the vectors $\nF_n$, see Section~\ref{sec:characteristic-quantities}. An example of this calculation is shown in Fig.~\ref{fig:plate_quantities}, where characteristic mode current distributions are reconstructed using FDTD simulation data. When compared to characteristic currents and characteristic far fields obtained using conventional EFIE MoM~\eqref{eq:CMAZ} (not shown) the results are visually indistinguishable.

\subsection{Structure Involving Dielectric and Metallic Elements}

\begin{figure}[t]
    \centering
    \includegraphics[width=\columnwidth]{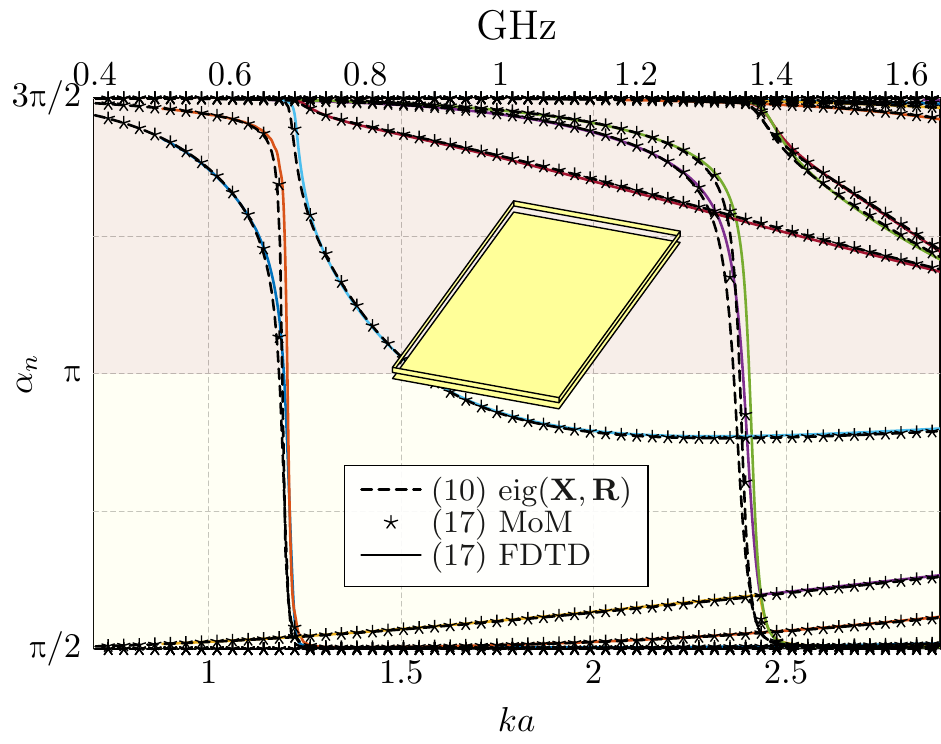}
    \caption{Characteristic angles~$\alpha_n$ of an air-filled PEC mobile device model. Data are shown for calculations using the system matrix formulation~\eqref{eq:CMAZ} (AToM, dashed lines) and the scattering dyadic formulation~\eqref{eq:cm-s-discrete} using MoM (FEKO,~$\star$, $N_\T{q} = 50$), and  FDTD (CST, solid lines, $N_\T{q} = 110$).}
    \label{fig:phone_alpha}
\end{figure}

\begin{figure}[t]
    \centering
    \includegraphics[width=\columnwidth]{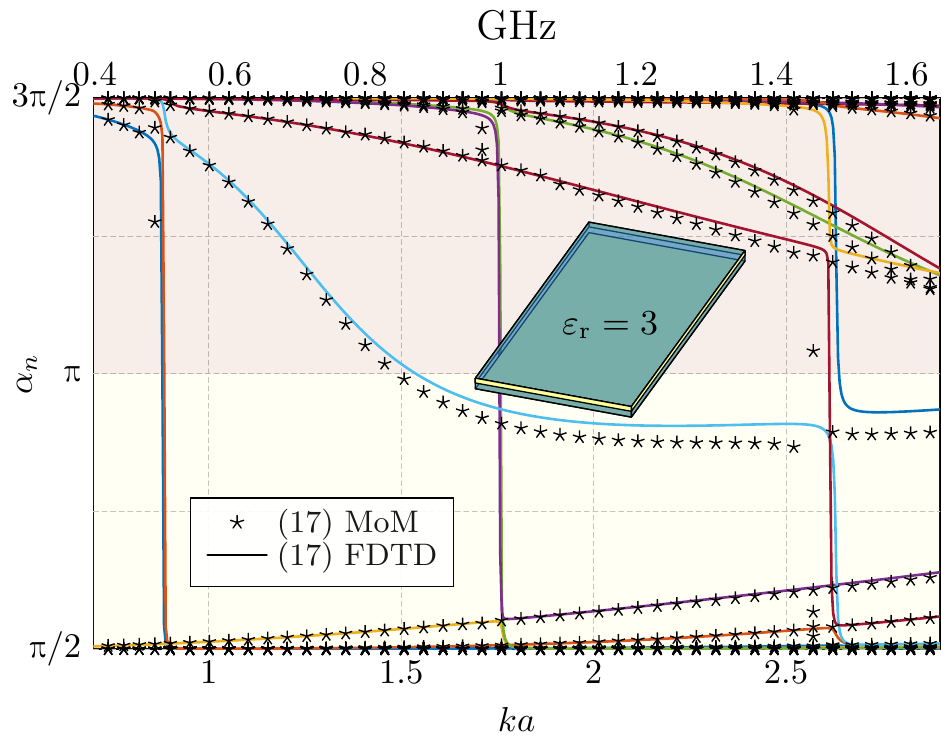}
    \caption{Characteristic angles~$\alpha_n$ of a dielectric-filled lossless PEC mobile device model. Data are shown for calculations using the scattering dyadic formulation~\eqref{eq:cm-s-discrete} using MoM (FEKO, surface equivalence,~$\star$, $N_\T{q} = 50$), and  FDTD (CST, solid lines, $N_\T{q} = 110$).}
    \label{fig:phone_diel_alpha}
\end{figure}

A significant advantage of the scattering dyadic formulation of characteristic modes is its applicability to structures containing arbitrary material distributions\footnote{Lossy materials can be introduced to characteristic mode calculations, but at the cost of losing far-field orthogonality and complication of the interpretation of modal quantities~\cite{Gustafsson_etal2021_CMAT_Part2}.  For simplicity, here we only consider lossless materials.}.  To demonstrate this feature, we consider an example representing a mobile device consisting of a rectangular PEC plate of dimension $150\times 75\,\T{mm}^2$ underneath a thin PEC rim following its perimeter.  The rim is $3.5\,\T{mm}$ tall and is separated from the rectangular plate by a gap of $2.5\,\T{mm}$.  The interior region of the device is filled with a lossless dielectric material with relative permittivity $\varepsilon_\T{r}$. Air-filled ($\varepsilon_\T{r} = 1$) and general dielectric-filled ($\varepsilon_\T{r} = 3$) versions of this structure are sketched in Figs.~\ref{fig:phone_alpha} and~\ref{fig:phone_diel_alpha}.

Characteristic modes of the air-filled structure were evaluated using~\eqref{eq:CMAZ} (AToM) and by~\eqref{eq:cm-s-discrete} with the scattering dyadic evaluated according to Sections~\ref{sec:MoM} using (FEKO) and~\ref{sec:FDTD} (CST). In terms of characteristic angles~$\alpha$, shown in Fig.~\ref{fig:phone_alpha}, the observed agreement between the methods is excellent. The dielectric-filled structure was analyzed using only the scattering dyadic method. Within FEKO, a surface equivalent treatment of the dielectric material was used for this calculation. Although no reference data are available in for this example, the observed agreement of characteristic angles from two different computational schemes gives assurance of the validity of the results. The overall effect of the dielectric filling is a lowering of sharp resonant frequencies by approximately~$30\,\%$ in~Fig.~\ref{fig:phone_diel_alpha} ($\left\{0.50\,\T{GHz}, 1.0\,\T{GHz}\right\}$) as compared to~Fig.~\ref{fig:phone_alpha} ($\left\{ 0.68\,\T{GHz}, 1.4\,\T{GHz} \right\}$). These resonances are predominantly governed by the current density on the rim and are therefore strongly affected by the presence of the dielectric.

\subsection{Multilayer Spheres}

\begin{figure}[t]
    \centering
    \includegraphics[width=\columnwidth]{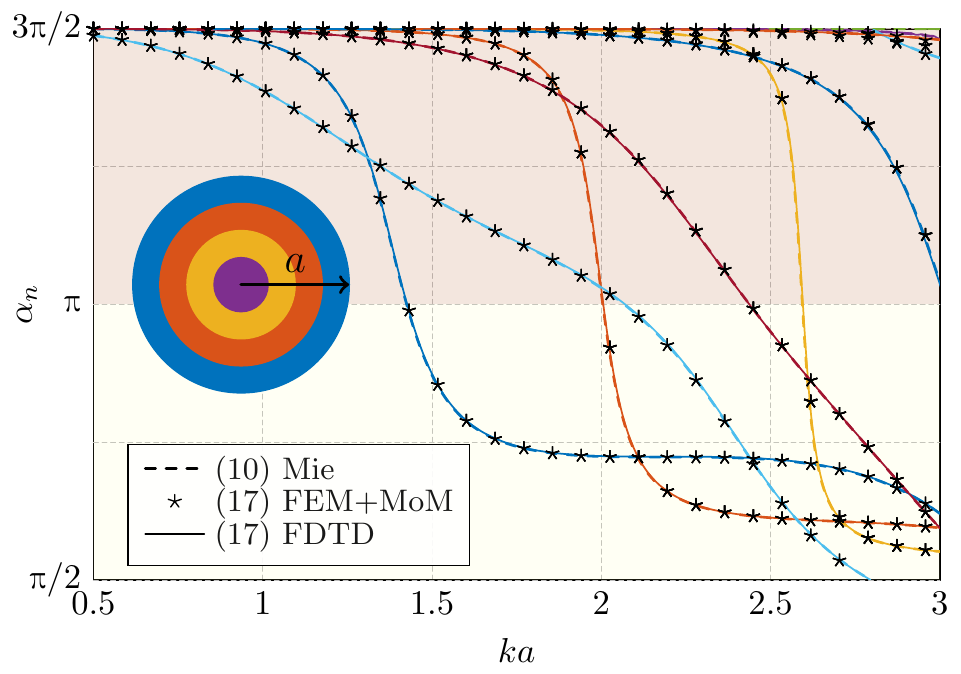}
    \caption{Characteristic angles~$\alpha_n$ of a multilayered dielectric sphere (illustrated in the top left of image). The layers have a thickness of~$0.25a$ and a relative permittivity of $\left\{3, 5, 8, 2\right\}$ (from the most central layer to the most outward). Data are shown for calculation using Mie series (dashed lines) and the scattering dyadic formulation~\eqref{eq:cm-s-discrete} using FDTD (CST, solid lines, $N_\T{q} = 74$) and FEM (FEKO,~$\star$, MoM used as radiation boundary condition, $N_\T{q} = 38$).}
    \label{fig:dielecsphere}
\end{figure}

\begin{figure}[t]
    \centering
    \includegraphics[width=\columnwidth]{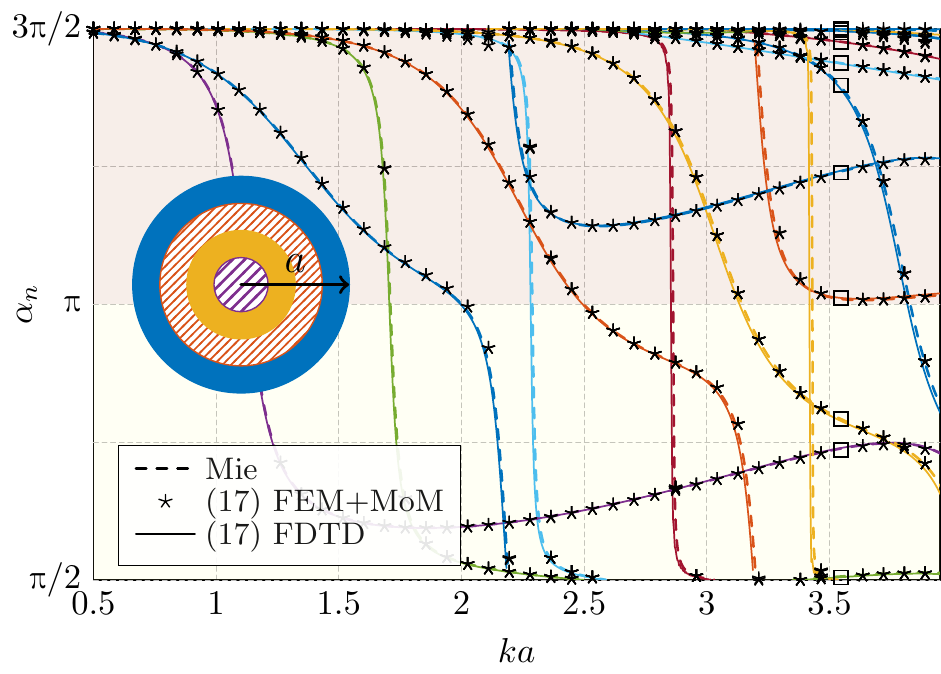}
    \caption{Characteristic angles~$\alpha_n$ of a multilayered dielectric-magnetic sphere (illustrated in the top left of image). The layers have a thickness of $0.25 a$ and a relative permittivity of $\left\{1, 5, 1, 2\right\}$ and the relative permeability of $\left\{3, 1, 8, 1\right\}$ (from the most central layer to the most outward). Data are shown for calculation using Mie series (dashed lines) and the scattering dyadic formulation~\eqref{eq:cm-s-discrete} using FDTD (CST, solid lines, $N_\T{q} = 74$) and FEM (FEKO,~$\star$, MoM used as radiation boundary condition, $N_\T{q} = 38$).}
    \label{fig:dielecmagnsphere}
\end{figure}

In order to obtain a comparison with analytical data, two spherical multilayer structures are studied. The first is a layered dielectric sphere consisting of four layers of thickness~$0.25 a$ with relative permittivities of~$\varepsilon_\T{r} = \left\{3, 5, 8, 2\right\}$ (inner layer to outer layer). The second structure contains layers with relative permittivities~$\varepsilon_\T{r} = \left\{1, 5, 1, 2\right\}$ and relative permeabilities~$\mu_\T{r} = \left\{3, 1, 8, 1\right\}$. Eigenangles~$\alpha_n$ of tracked modes for the dielectric and dielectric / magnetic spheres are depicted in Figs.~\ref{fig:dielecsphere} and~\ref{fig:dielecmagnsphere}, respectively. The dashed lines correspond to analytical Mie series solutions\footnote{The diagonal transition matrix of layered spheres may be constructed analytically using the Mie series~\cite{Kristensson_ScatteringBook} and this matrix may, in turn, be used to obtain characteristic values via~\eqref{eq:TCMA}.}, solid lines are from FDTD simulations in CST Studio Suite, and the markers are from FEM simulation in Altair~FEKO. The square markers at $ka \approx 3.54$ were evaluated 10\,\% off the position of the equidistant frequency point ($ka \approx 3.55$) due to the instability of FEKO's solver at the original frequency.

Within the results for the dielectric/magnetic layered sphere, features of note include significantly lower modal resonance frequencies compared to the dielectric layered sphere and a wide-band resonance near~$ka = 3.5$. The fact that characteristic modes for these general structures are known analytically means that these examples validate the accuracy of the methods proposed in this work and can serve as benchmark problems for other characteristic mode calculation techniques that may be developed in the future.

\section{Discussion}
\label{sec:disc}

\begin{figure}[t]
    \centering
    \includegraphics[width=\columnwidth]{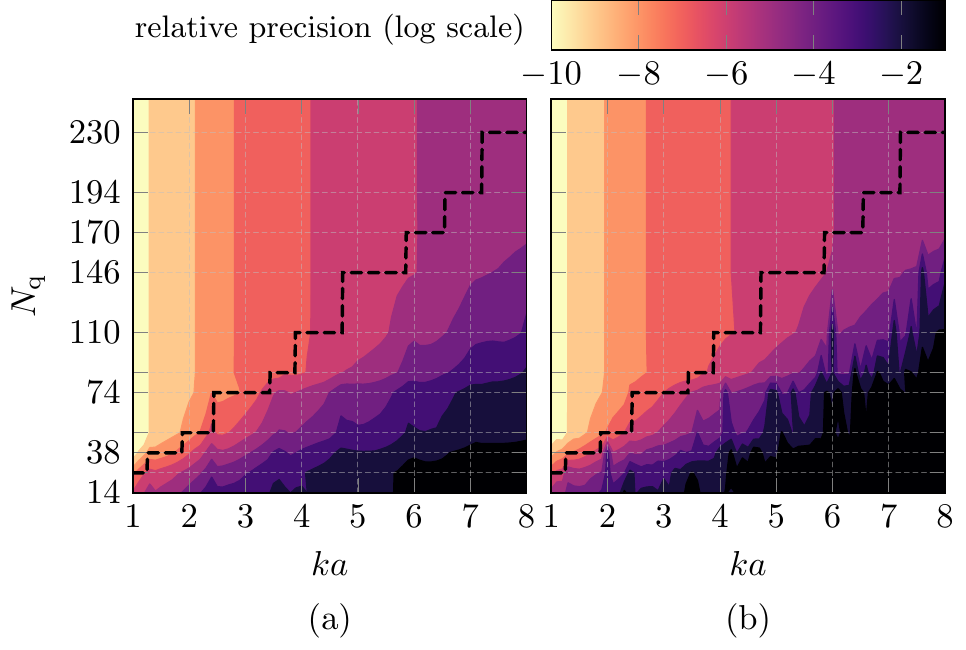}
    \caption{The panes show average error in magnitude (a) and phase (b) of the eigenvalues~$t_n$. First 25~eigenvalues (ordered with respect to decreasing modal significance) are evaluated using scattering dyadic formulation~\eqref{eq:cm-s-discrete} and system matrix decomposition~\eqref{eq:CMAZ}. The eigenvalues $t_n$ were first converted to eigenvalues~$s_n = 2 t_n + 1$, \cite{Gustafsson_etal2021_CMAT_Part1}, and the error in absolute value was evaluated as $(1/N)\sum_n |s_n|-1$ and in phase as $(1/N)\sum_n \min \left\{2\pi-\Delta_n, \Delta_n\right\}$ with $\Delta_n = \angle s_n^\T{MoM} - \angle s_n$ and $s_n^\T{MoM}$ representing data evaluated via~\eqref{eq:CMAZ}. Both evaluations~\eqref{eq:CMAZ} and~\eqref{eq:cm-s-discrete} are dependent on system matrix, while only~\eqref{eq:cm-s-discrete} depends on degree of the quadrature~$N_\T{q}$. Therefore, data from~\eqref{eq:CMAZ} serve as a reference. The estimate~\eqref{eq:LebQuadEstimate} is depicted as the solid dashed lines. The values of $N_\T{q}$ are taken only from the set of allowed Lebedev quadrature degrees. The model used in this calculation is the metallic rim from Fig.~\ref{fig:phone_alpha} discretized with $1212$~basis functions and with the ground plane removed from one of the quadrants (to enrich the characteristic mode spectrum and to break the symmetries).}
    \label{fig:precision}
\end{figure}

The scattering dyadic method proposed in this paper has several notable advantages over the classical characteristic mode decomposition based on system (impedance) matrices.  First, the matrix decomposition is a standard eigenvalue problem, as opposed to the generalized eigenvalue problem utilized in the impedance-based decomposition~\cite{HarringtonMautz_TheoryOfCharacteristicModesForConductingBodies}. Second, the rank of the scattering dyadic matrix (\ie{}, the number of calculated modes) can easily be controlled by the selected number of quadrature points. The type of the quadrature and its degree controls the precision, see Fig.~\ref{fig:precision}. In this respect, the Lebedev quadrature~\cite{LebedevLaikov_QuadratureRuleSphere} excels in number of points required and in knowledge of how many spherical waves are taken into account. As an alternative to Lebedev quadrature with the estimate~\eqref{eq:LebQuadEstimate}, other quadrature rules enabling reuse of integration points, \eg{},~\cite{Behjoo2021}, would, however, allow for iterative decomposition based on checking for predefined error in, \eg{}, modal significance. These computational aspects accelerate characteristic mode decomposition.

An additional benefit of the scattering dyadic approach is that, similarly to~\eqref{eq:TCMA}, the eigenvectors are themselves the characteristic far fields, used often for tracking the eigenvalues and eigenvectors~\cite{SafinManteuffel_AdvancedEigenvalueTrackingofCM, Gustafsson_etal2021_CMAT_Part2}. Therefore, it is not necessary to evaluate characteristic far fields from surface current densities, as is required in the classical impedance decomposition. The advanced tracking routines based on scattering matrix interpolation and rational fitting~\cite{Gustafsson_etal2021_CMAT_Part2} can be used to mitigate the computational burden of frequency-domain solvers.

A unique feature is a principal possibility to decompose datasets provided by a bistatic measurement (including phase) \cite{Geffrin2009} or datasets generated by co-simulation of full-wave and circuit designers. Other possibilities include investigation of objects above infinite ground planes (filled by stratified media), periodic structures, or specific scenarios addressed with Inagaki modes~\cite{InagakiGarbacz_EigenfunctionsOfCompositeHermitianOperatorsWithApplicationToDiscreteAndContunuousRadiatingSystems}.

\section{Conclusion}
\label{sec:concl}

In this work, the scattering dyadic is used to define characteristic mode decomposition. The final formulation is independent of the choice of electromagnetic solver and it can easily be implemented in many academic and commercial software packages as it requires only the solution of plane wave scattering problems. Wrappers for MoM, FEM, and FDTD in several commercial software packages are available as supplementary material~\cite{ScatDyad22_SuplMat}.

The method is general and valid for arbitrary materials of arbitrary distribution. Lossy materials can be treated as well, however this comes at the cost of losing modal far-field orthogonality. 

The eigenvectors produced by the scattering dyadic method simultaneously represent the characteristic far fields as well as characteristic excitations. Other characteristic quantities often used for antenna design, such as modal surface currents, can be reconstructed by exciting the obstacle with the selected characteristic excitation. This also implies that other non-standard modal quantities, such as modal near fields, may be directly calculated using the same process, which may be advantageous in systems involving inhomogeneous dielectrics.

Calculation of characteristic modes using the scattering dyadic may be accelerated in several ways. The advanced tracking algorithms utilizing interpolation and fitting can be used in the frequency domain to significantly reduce the number of frequency points required to reconstruct modal data over broad bandwidths. In the time domain, the entire frequency sweep can be recovered from a single simulation through the proper choice of the excitation signal. Finally, as the number of plane waves used to construct the scattering dyadic grows, an adaptive quadrature rule reusing quadrature points from the previous iteration may provide significant computational speed-up.

The proposed approach also opens up several future areas of work in characteristic modes. Use of the scattering dyadic makes it possible to obtain characteristic data from co-simulation of full-wave and linear circuit designs, since this framework only demands an input-output relation between impinging plane waves and scattered far fields.  Additionally, the scattering dyadic can, in principle, be constructed using bi-static scattering data, allowing for the reconstruction of characteristic numbers and far fields from laboratory measurements. Other open topics, such as the characteristic modes of patch antennas over layered media, infinite ground planes, or periodic environments are also possible to explore with the scattering dyadic methodology.

\appendices


\section{Equivalence of T-matrix and scattering dyadic eigenvalue problems}
\label{app:TtoS}

Alternatively to the description via scattering dyadic~\eqref{eq:scattering-dyadic-def} and plane waves, the incident field can be represented as a sum of spherical vector waves 
\begin{equation}
     \V{E}^\T{i}(\V{r}) = k\sqrt{\ZVAC}\sum_\alpha a_{\alpha}\,\M{u}_\alpha^{(1)}(k\V{r})
     \label{eq:ei-regular}
\end{equation}
and the scattered far field as a sum of vector spherical harmonics 
\begin{equation}
     \V{F}(\hat{\V{r}}) = \sqrt{\ZVAC} \sum_\alpha  f_{\alpha}\bfSW_\alpha (\hat{\V{r}}),
     \label{eq:f-exp-def}
\end{equation}
where 
\begin{equation}
\bfSW_\alpha (\UV{r}) = - \left( -\J \right)^{-l + \tau} \V{A}_\alpha(\UV{r}),
\end{equation}
see~\cite[Appendix~C]{Kristensson_ScatteringBook} for details.

Under these expansions, the object's scattering characteristics are represented by a transition matrix $\M{T}$ mapping incident (regular) wave coefficients to scattered (outward) wave coefficients via
\begin{equation}
    \M{f} = \M{T}\M{a}.
    \label{eq:tmat-def}
\end{equation}
It is this transition matrix that forms the definition of characteristic modes originally proposed by Garbacz~\cite{Garbacz_TCMdissertation}
\begin{equation}
\M{T}\M{f}_n = t_n\M{f}_n,
\label{eq:cm-t}
\end{equation}
where, by \eqref{eq:f-exp-def}, \eqref{eq:ei-regular}, and \eqref{eq:tmat-def}, the eigenvectors $\M{f}_n$ represent both characteristic incident and scattered fields.

The T-matrix may be represented in terms of the scattering dyadic~$\V{S}(\UV{r},\UV{r}')$ via~\cite{Kristensson_ScatteringBook}
\begin{equation}
T_{\alpha \beta} = \int \limits_{4 \pi} \int \limits_{4 \pi} \bfSW_\alpha^\ast (\UV{r}) \cdot \V{S}(\UV{r},\UV{r}') \cdot \bfSW_\beta (\UV{r}') \D{\varOmega}\D{\varOmega}'.
\label{eq:tab-s}
\end{equation}
Using relations \eqref{eq:f-exp-def} and \eqref{eq:tab-s}, the eigenvalue problem in \eqref{eq:cm-t} may be written as 
\begin{equation}
    \frac{1}{\sqrt{\ZVAC}}\int \limits_{4 \pi} \int \limits_{4 \pi} \bfSW_\alpha^*(\UV{r})\cdot\V{S}(\UV{r},\UV{r}')\cdot\V{F}_n(\UV{r}')\D{\varOmega}\D{\varOmega'} = t_n f_{n,\alpha}
    \label{eq:app-a-1}
\end{equation}
for all indices $\alpha$. Because of the orthogonality of the chosen basis~\cite[Appendix~C-4]{Kristensson_ScatteringBook},
\begin{equation}
    \int \limits_{4 \pi} \bfSW_\alpha^*(\UV{r}) \cdot \bfSW_\beta(\UV{r})\,\D{\varOmega} = \delta_{\alpha\beta}, 
\end{equation}
the right hand side of \eqref{eq:app-a-1} may be rewritten using \eqref{eq:f-exp-def} to give 
\begin{multline}
\int \limits_{4 \pi} \bfSW_\alpha^*(\UV{r})\cdot\int \limits_{4 \pi} \V{S}(\UV{r},\UV{r}') \cdot \V{F}_n(\UV{r}')\D{\varOmega'}\D{\varOmega} \\ = t_n \int \limits_{4 \pi} \bfSW_\alpha^*(\UV{r})\cdot \V{F}_n(\UV{r})\D{\varOmega}\quad \text{for all } \alpha.
\end{multline}
The basis $\{\bfSW_\alpha\}$ is complete for vector functions over the unit sphere~\cite[Appendix~C-4]{Kristensson_ScatteringBook}, which implies equality of the integrands in the above expression, \ie{},
\begin{equation}
    \int \limits_{4 \pi} \V{S}(\UV{r},\UV{r}')\cdot\V{F}_n(\UV{r}')\D{\varOmega'} = t_n\V{F}_n(\UV{r}).
\end{equation}

\section{Relation of Characteristic Modes and Impedance Matrix}
\label{app:StoZ}

The conversion of eigenvalue equation~\eqref{eq:cm-s} into a matrix eigenvalue problem can also be done using method of moments applied to electric field integral equation~\cite{Harrington_FieldComputationByMoM}. Within that paradigm, the interaction of the scatterer with incident field is described as
\begin{equation}
\label{eq:appC:ZIV}
\M{Z} \M{I} = \M{V},
\end{equation}
where~$\M{Z}$ is the impedance matrix,~$\M{I}$ is a column vector of current expansion coefficients 
and column vector~$\M{V}$ is a collection of terms
\begin{equation}
V_i = \int \limits_{\mathbb{R}^3} \bfMoM_i \left( \V{r} \right) \cdot \V{E}^\T{i} \left( \V{r} \right) \D{V}
\end{equation}
projecting incident electric field onto the basis functions~$\bfMoM_i$ expanding the current density.

In this paradigm the~$\gamma$-th polarization component of far field in direction~$\hat{\V{r}}$ is given as~$F_\gamma \left(\hat{\V{r}} \right) = \FFop_\gamma \left(\hat{\V{r}} \right) \M{I}$, where
$\FFop_\gamma$ is a row vector with components
\begin{equation}
K_{\gamma,i}(\UV{r}) = -\J \dfrac{\ZVAC k}{ 4 \pi} \int_{\mathbb{R}^3}  \bfMoM_i \left( \V{r}_1 \right) \cdot \hat{\V{\gamma}} \T{e}^{\J k \hat{\V{r}} \cdot \V{r}_1} \D{V_1},
\label{eq:ffop-def}
\end{equation}
where~$\hat{\V{\gamma}}$ denotes the unit vector along the~$\gamma$-polarization component.

Assuming, that incident field is a plane wave with unit amplitude, polarized along vector~$\hat{\V{\gamma}}$ and propagating in direction~$\hat{\V{r}}'$ 
\begin{equation}
\V{E}^\T{i} \left( \V{r} \right) = \UV{\gamma} \T{e}^{- \J k \hat{\V{r}}' \cdot \V{r}},
\label{eq:ei-pw}
\end{equation}
it can be seen that
\begin{equation}
\label{eq:appC:FV}
\M{V}_\gamma \left(\hat{\V{r}}' \right) = -\J \dfrac{4 \pi}{\ZVAC k} \FFop_\gamma^\herm \left(\hat{\V{r}}' \right).
\end{equation}
Relation~\eqref{eq:appC:FV} together with~\eqref{eq:appC:ZIV} and~\eqref{eq:scattering-dyadic-def} can be used to evaluate components of scattering  dyadic as
\begin{equation}
\label{eq:appc:SFZF}
S_{\gamma\gamma'} \left(\hat{\V{r}},\hat{\V{r}}' \right) = -\dfrac{1}{\ZVAC} \FFop_\gamma \left(\hat{\V{r}} \right) \M{Z}^{-1}  \FFop_{\gamma'}^\herm \left(\hat{\V{r}}' \right),
\end{equation}
where invertibility of the matrix $\M{Z}$ is assumed.  

To demonstrate the equivalence between the eigenvalue problems \eqref{eq:cm-s} and \eqref{eq:CMAZ}, consider a characteristic mode calculated from \eqref{eq:cm-s} defined by a modal far-field $\V{F}_n$. This modal far field can be written in terms of a method of moments representation of the modal current $\M{I}_n$ as
\begin{equation}
    F_{\gamma,n} \left(\hat{\V{r}} \right) = \FFop_\gamma \left(\hat{\V{r}} \right) \M{I}_n.
    \label{eq:cm-ff}
\end{equation}
Similarly to the construction of single plane wave excitations in \eqref{eq:ei-pw} and \eqref{eq:appC:FV}, the relations in \eqref{eq:cm-en}, \eqref{eq:VnFromEnI}, and \eqref{eq:ffop-def} can be combined to write the method of moments representation of the characteristic excitation for this mode as
\begin{multline}
    \M{V}_n = -\frac{1}{Z_0 t_n}\sum_\gamma \int\limits_{4 \pi} \FFop_\gamma^\T{H}(\UV{r}') F_{\gamma,n}(\UV{r}')\D{\varOmega'}\\
    = -\frac{1}{Z_0 t_n}\sum_\gamma \int\limits_{4 \pi} \FFop_\gamma^\T{H}(\UV{r}') \FFop_\gamma \left(\hat{\V{r}}' \right) \M{I}_n\D{\varOmega'},
    \label{eq:cm-equiv-a}
\end{multline}
where the final right-hand side is obtained by substitution of \eqref{eq:cm-ff}.  For lossless structures, the integrated power radiated in all directions corresponds to the total real power dissipated by the current, i.e.,
\begin{equation}
    \M{I}^\herm \dfrac{1}{Z_0} \sum_\gamma \int \limits_{4 \pi} \FFop_\gamma^\herm \left(\hat{\V{r}} \right) \FFop_\gamma \left(\hat{\V{r}} \right) \D{\varOmega} \, \M{I} =  \M{I}^\herm \M{R} \M{I}.
\end{equation}
Because the above relation holds for any current $\M{I}$, we have
\begin{equation}
    \dfrac{1}{Z_0} \sum_\gamma \int \limits_{4 \pi} \FFop_\gamma^\herm \left(\hat{\V{r}} \right) \FFop_\gamma \left(\hat{\V{r}} \right) \D{\varOmega} = \M{R}.
\end{equation}
Substituting this identity into \eqref{eq:cm-equiv-a} leads to
\begin{equation}
    \M{V}_n = -\frac{1}{t_n}\M{R}\M{I}_n,
\end{equation}
which, when combined with \eqref{eq:appC:ZIV} gives
\begin{equation}
    \M{Z}\M{I}_n = -\frac{1}{t_n}\M{R}\M{I}_n.
\end{equation}
Rearranging and substitution of \eqref{eq:tn2lamn} yields the typical characteristic mode eigenvalue problem in \eqref{eq:CMAZ}.


\begin{IEEEbiography}[{\includegraphics[width=1in,height=1.25in,clip,keepaspectratio]{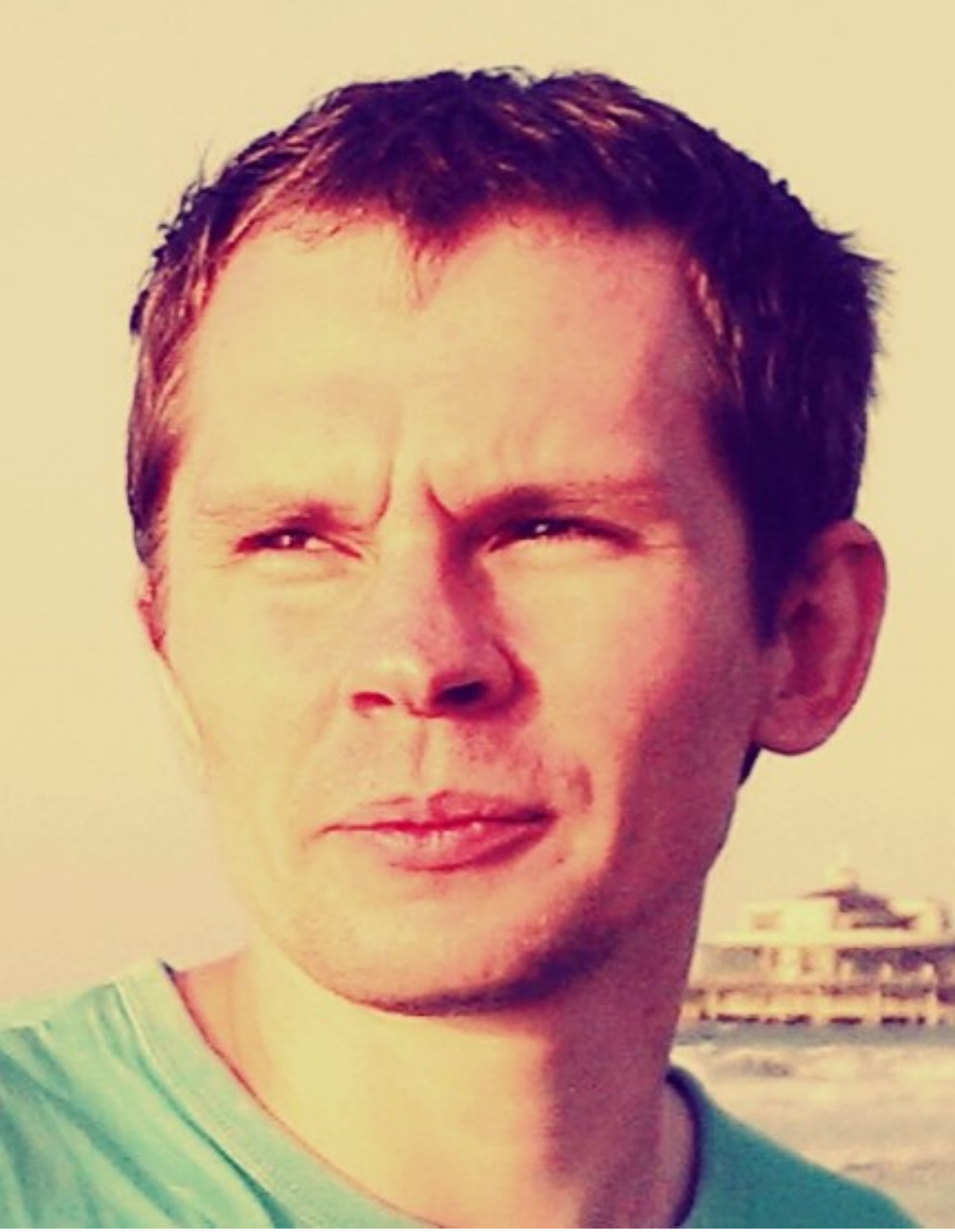}}]{Miloslav Capek}
(M'14, SM'17) received the M.Sc. degree in Electrical Engineering 2009, the Ph.D. degree in 2014, and was appointed Associate Professor in 2017, all from the Czech Technical University in Prague, Czech Republic.
	
He leads the development of the AToM (Antenna Toolbox for Matlab) package. His research interests are in the area of electromagnetic theory, electrically small antennas, antenna design, numerical techniques, and optimization. He authored or co-authored over 100~journal and conference papers.
\end{IEEEbiography}

\begin{IEEEbiography}[{\includegraphics[width=1in,height=1.25in,clip,keepaspectratio]{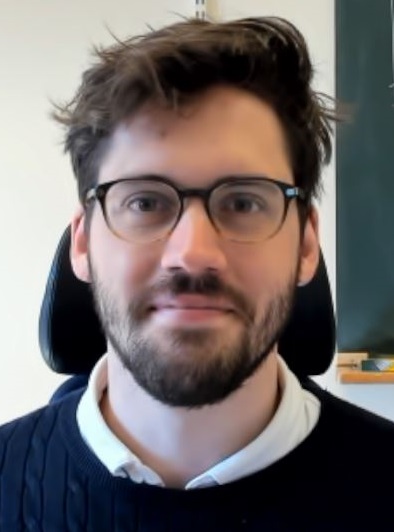}}]{Johan Lundgren}
(M'22) is a postdoctoral researcher at Lund University. he received his M.Sc degree in engineering physics 2016 and Ph.D. degree in Electromagnetic Theory in 2021 all from Lund University, Sweden. 

His research interests are in electromagnetic scattering, wave propagation, computational electromagnetics, functional structures, meta-materials, inverse scattering problems, imaging, and measurement techniques.

\end{IEEEbiography}

\begin{IEEEbiography}[{\includegraphics[width=1in,height=1.25in,clip,keepaspectratio]{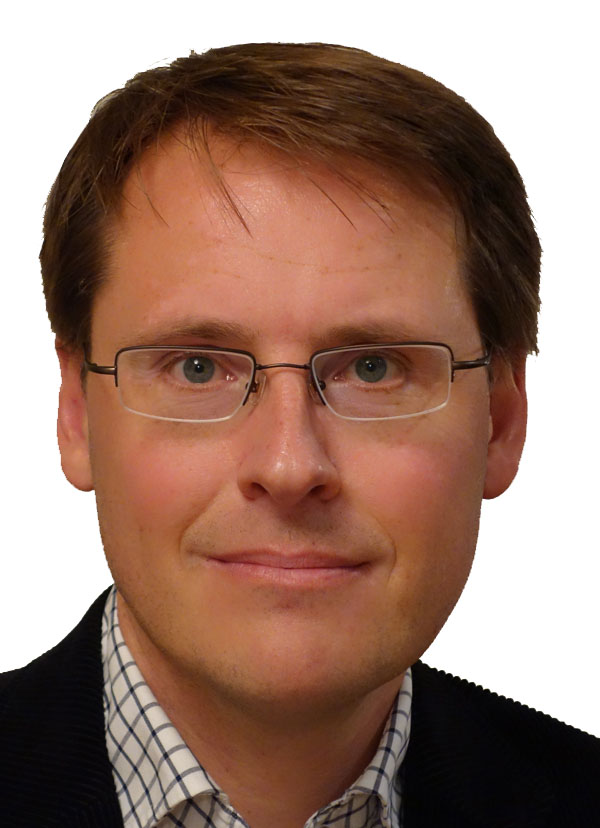}}]{Mats Gustafsson}
received the M.Sc. degree in Engineering Physics 1994, the Ph.D. degree in Electromagnetic Theory 2000, was appointed Docent 2005, and Professor of Electromagnetic Theory 2011, all from Lund University, Sweden.

He co-founded the company Phase holographic imaging AB in 2004. His research interests are in scattering and antenna theory and inverse scattering and imaging. He has written over 100 peer reviewed journal papers and over 100 conference papers. Prof. Gustafsson received the IEEE Schelkunoff Transactions Prize Paper Award 2010, the IEEE Uslenghi Letters Prize Paper Award 2019, and best paper awards at EuCAP 2007 and 2013. He served as an IEEE AP-S Distinguished Lecturer for 2013-15.
\end{IEEEbiography}

\begin{IEEEbiography}[{\includegraphics[width=1in,height=1.25in,clip,keepaspectratio]{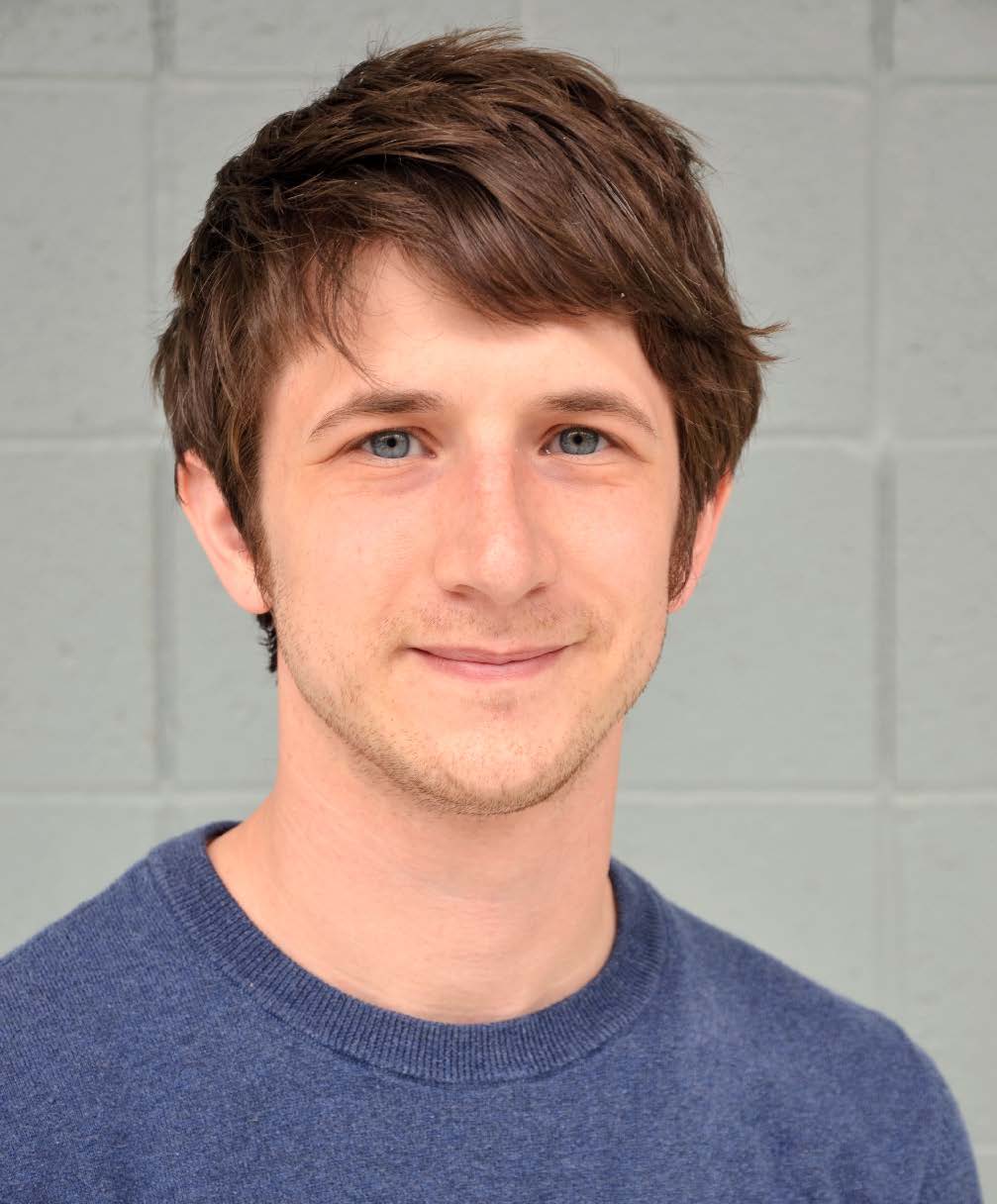}}]{Kurt Schab}
(S'09, M'16) is an Assistant Professor of Electrical Engineering at Santa Clara University, Santa Clara, CA USA. He received the B.S. degree in electrical engineering and physics from Portland State University in 2011 and the M.S. and Ph.D. degrees in electrical engineering from the University of Illinois at Urbana-Champaign in 2013 and 2016, respectively.  From 2016 to 2018 he was a Postdoctoral Research Scholar at North Carolina State University in Raleigh, North Carolina.  His research focuses on the intersection of numerical methods, electromagnetic theory, and antenna design.  
\end{IEEEbiography}

\begin{IEEEbiography}[{\includegraphics[width=1in,height=1.25in,clip,keepaspectratio]{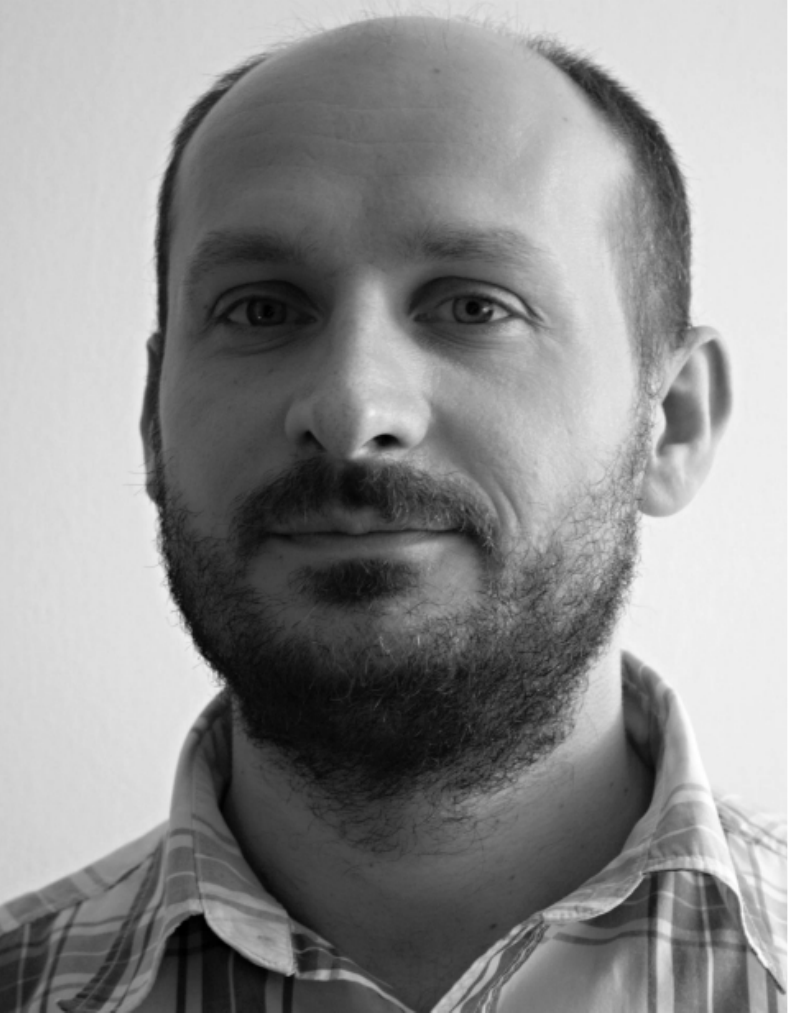}}]{Lukas Jelinek}
received his Ph.D. degree from the Czech Technical University in Prague, Czech Republic, in 2006. In 2015 he was appointed Associate Professor at the Department of Electromagnetic Field at the same university.

His research interests include wave propagation in complex media, electromagnetic field theory, metamaterials, numerical techniques, and optimization.
\end{IEEEbiography}

\end{document}